# Software Architecture

Author: Andre Adrian

Version: 23jul2005

## Table of Contents

## Introduction

What is Software Architecture? The rules, paradigmen, pattern that help to construct, build and test a serious piece of software[1]. It is the practical experience boiled down to abstract level. Software Architecture started with "Stepwise refinement"[2] from Niklaus Wirth and "Programming in the large"[3] by DeRemer and Kron or even before. It builds on System Engineering and the scientific method as established by Galileo Galilei: Measure what you can and make measureable what you can not. The experiment is more important then the deduction[4].

Pieces of information about software architecture are all over the internet. This paper uses citation as much as possible. The aim is to bring together the pieces, not to rephrase the wording.

As always, truth is out of the reach for man. But like the mathematicains in integral calculus we can create series of better and better approximations[5].

The author does not want to sell CASE tools to you - there are others in the marketplace to do this. I want to sell you some ideas. Ideas you can use for every software you write in whatever programming language you use. I want to give you the big picture. After 20 years in computer science I say: tools are nice but understanding is better.

Drawing the big picture is not only the topic of software architecture. Unified Modeling Language[6] has the same goal. But architecture is a broader term then language. Some critics even call the 13 different UML diagram types bloating the topic.

---

[1] http://www.bredemeyer.com/whatis.htm
[2] http://www.acm.org/classics/dec95/
[3] http://www.mapfree.com/sbf/tips/smalarg.html#deremer76
[4] http://www.rit.edu/~flwstv/galileo.html
[5] http://en.wikipedia.org/wiki/Approximation
[6] http://de.wikipedia.org/wiki/Unified_Modeling_Language
http://en.wikipedia.org/wiki/Unified_Modeling_Language

# Spiral model

http://en.wikipedia.org/wiki/Spiral_model

The spiral model is a development model combining elements of both design and prototyping-in-stages, in an effort to combine advantages of top-down and bottom-up concepts.

The spiral model was defined by Barry Boehm in his article A Spiral Model of Software Development and Enhancement from 1986. This model was not the first model to discuss iteration, but it was the first model to explain why the iteration matters.

# Hierarchical Decomposition

For most readers Structured Design (SD) is an outdated design method from the 1970s[7] promoted by Yourdon and Constantine. But SD promoted hierarchical decomposition, a very powerful idea that works fine together with the object oriented paradigma. Hierarchical Decomposition was also part of HIPO (Hierarchical Input Processing Output) - and is therefore even older then SD.

The important fact is: Hierarchical Decomposition is very valueable, but the rest of HIPO and SD are rubble today. The OO paradigma fills now this part.

A large system must be divided into manageable sub-systems (or components or modules or objects) in several layers. Hierarchical Decomposition uses a simple but powerful notation for this purpose. First the system is divided into top-level sub-systems. Then every top-level sub-system is divided into second-level sub-systems. The sub-systems are identified with a hierarchical numbering. A sub-system named 2. is a first-level sub-system, a sub-system named 2.3.1. is a third-level sub-system.

Upper level sub-systems are no objects. The lowest level sub-systems should be designed as objects. Without Hierarchical Decomposition it is more difficult to "objectify" a system.

Normally sub-systems are designed with the hardware devices in mind. For example mobile phone (user terminal) and base station are two sub-systems in a cellular telephone system.

The hierachical decomposition allows to go into different depth of sub-system design at different times in the design without loosing the big picture of the full system. The upper-level sub-system contains all details of the included lower-level sub-systems. This is important for the input and output of sub-systems. Normally in hierarchical decomposition the input and output of upper-level sub-systems have to get updated (enhanced) due to new needs that came into the designers focus while working on lower-level sub-systems.

---

[7] http://www.win.tue.nl/~wstomv/quotes/structured-design.html

# Bottom-up development

The bottom-up development (also called agile software development or extreme programming) is the spiral model used on the algorithms of the system. The bottom-up development is the opposite of the top-down hierarchical decomposition. You should use hierarchical decomposition to create your master plan, but you should use bottom-up development to drive one development cycle.

Depending on the system, algorithms are the easiest or the hardest part of all. Easy if you can use cook-book algorithms, hard if you have to develop algorithms. To develop an algorithm the iterative bottom-up approach is advised: You implement the most important sub-system from your hierarchical decomposition and then you enhance this partial system into the complete system in several cycles. It is important to remember that the partial system is not a throw-away prototype. In the iterative bottom-up method you have to use the same programming language and development environment that you use for the final system. The goal is not to build mock-ups, the goal is to bring in early testing and early user-feedback into the development.
Because of the cycle rate of weeks or even days, the user is much more involved then before. Instead of waiting 2 years for the software team to make the first release, every cycle the user has to audit the progress. And because the software developer, tester and manager did learn something new in the last cycle the planning of the next cycle has to get adapted to this new knowledge.
Last cycle insight changes normally the sub-systems in the hierarchical decomposition and ripples through all sub-system levels into changes of the master plan, the top-level system plan.
From a team-building point of view it makes sense to bring all participants (user, developer, tester) into one team. But because of social dynamics this approach has drawbacks: the heterogenous team members can compromise within themselfs on a team-view about the system that is not in line with the requested system behaviour that is defined by the expectations of all users, not only the users in the team.
To make the audit not a farce, the audit after every cycle has to involve user and tester participants with the necessary knowledge that are outsite the heterogenous team.

http://www.dmreview.com/article_sub.cfm?articleId=1025869

You will likely have some management challenges as you execute your cycles, here are some things to watch out for:

- Delaying a cycle due to some issue. Never delay a cycle because of some act of discovery. Push the issue into some future cycle or remove it from the release altogether. Our goal here is to create periodic points where we can access our development state. If you delay, your state is unknown.
- Too much change. Watch your rate of introduction of change, especially in the development cycles (it's assumed in the stabilization cycles). Data model changes are a good example. Major schema rework just results in the same work being performed over and over again. Create "bliss" cycles where developers can focus just on moving forward, even if the data model has a few flaws. Remember the goal of any development task is to make the "unknown, known," and you need to move forward to discover what else you don't know, or it will surprise you later.

# Model View Controller

http://ootips.org/mvc-pattern.html

The MVC paradigm is a way of breaking an application, or even just a piece of an application's interface, into three parts: the model, the view, and the controller. MVC was originally developed to map the traditional input, processing, output roles into the GUI realm:

```
 Input --> Processing --> Output
 Controller --> Model --> View
```

http://www.cs.indiana.edu/~cbaray/projects/mvc.html

The MVC pattern is surprisingly simple, yet incredibly useful. Essentially, the pattern forces one to think of the application in terms of these three modules -

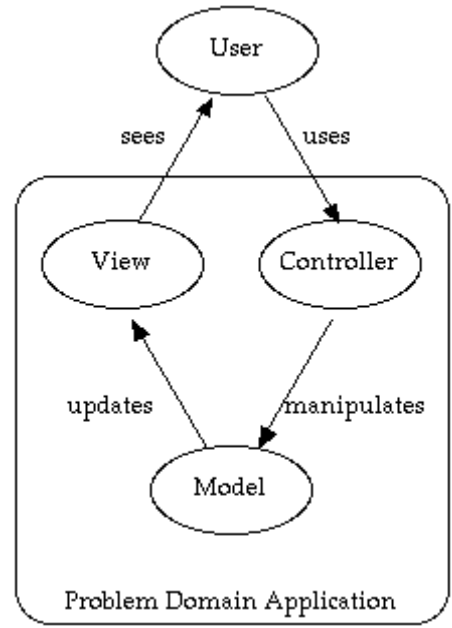

- Model : The core of the application. This maintains the state and data that the application represents. When significant changes occur in the model, it updates all of its views
- Controller : The user interface presented to the user to manipulate the application.
- View : The user interface which displays information about the model to the user. Any object that needs information about the model needs to be a registered view with the model.

http://wiki.tcl.tk/6225

Model-View-Controller [1] (often abbreviated "MVC") is a way of working with GUIs that provide access to complex data, and it seems to scale really quite well indeed. You do it by separating the code to implement the Model (your underlying data is, say, an updatable list or a read-only tree) from the View (how your data looks when drawn on the screen) and the Controller (how your model responds to updates to the view) though often the view and controller are joined together.

Some features of Tk operate in this way; notably the -variable option to entry, the -textvariable option to label, button et al, and the -listvariable option to listbox.

It's wise to design the rest of your interface that way as well. The trace command (particularly [trace variable]) is very useful for implementing triggers that update a view when the underlying model changes. The bindtags command is useful for associating a widget with one or more controllers other than the standard ones that Tk provides; occasionally, you also need to resort to overloading a widget command. For a worked example, see the page entitled, "Text variable for text widget". In that example, the widget command is overloaded to notify the model when the user attempts to change the text contents. A trace notifies the view when the model is updated. A bind tag is also used so that everything gets cleaned up correctly when the window is destroyed.

http://wiki.tcl.tk/9219

if 0 {Richard Suchenwirth 2003-06-27 - As yet another attempt to teach programming to children, this tiny plaything (both in length of code and screen estate) does multiplication on the fly for the factors you enter into the two entries.

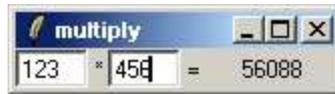

All communication happens via textvariables and traces. I'm not sure whether this already counts as a minuscule Model / View / Controller example:

the variables a,b,c are the models
the label .c is a view on the result, c
the entries .a, .b are controllers for the user to change the model

But in any case, it's a pretty minimal example '(also for fancy widget names - I didn't believe they could be called .* or .= before I tried ...) }

```
proc main {} {
    global a b c
    entry .a -width 5 -textvariable a
    label .* -text *
    entry .b -width 5 -textvariable b
    label .= -text =
    label .c -width 10 -textvariable c
    eval pack [winfo children .] -side left

    foreach factor {a b} {
        trace variable $factor w {recompute}
    }
}
```

if 0 {In this proc which tries to recompute (that might fail if an entry has been cleared), we ignore the three arguments that a trace gets automatically added - we know what we want:}

```
proc recompute {- - -} {
    global a b c
    catch {set c [expr {$a * $b}]}
}
```

# Finally, let's go (it's a nice pattern to start with proc main, and end with calling it..)

```
main
```

# Callback

The topic of callback (event handler, software interrupt) is important for GUI and Real-Time programming. The connection between controller to model in MVC is a callback, the connection between model and view can be a normal function call.

Callbacks are used to connect the world outside the program with the program. The events are asynchronous. The program expects these events, but the order of events or the exact time the event will happen are not known before.

First action is to register the callback (install the event handler). The run-time environment of the program stores the details of the callback. If the condition (trigger) of the callback is fulfilled, the callback is called. The run-time environment for GUI-programs are typical libs like the Xt (X Windows System Toolkit) or the Tcl/Tk run-time lib. For RT-programs the POSIX system call select() is normally the callback register and call function. The operating system is the run-time environment.

GUI and RT programming can be seen as writing extensions to the GUI or RT run-time. The extensions are the callbacks. The main control loop of the program is in the run-time environment, not in the program itself.

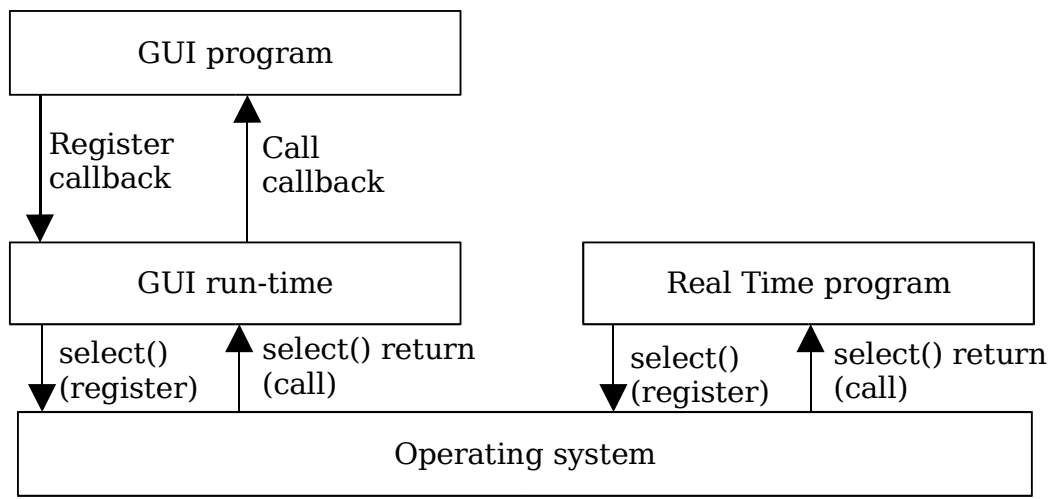

The callback triggers in Tcl/Tk can be user GUI activity, timeout, network receive and variable access. The examples are mostly from the intercom.tcl source code.

```
# Register user GUI activity (button press release)
# The callback function keyRelease is called with the parameter t1
.t1 configure -command [list keyRelease t1]
```

```
# register timeout
# The callback function checkping is called with no parameter
after 200 [list checkping]
```

```
set sock [socket 127.0.0.1 4999]        ;# open TCP connection, client side
# register network receive
# The callback function recv is called with no parameter
fileevent $sock readable recv
```

```
# register variable write
# The callback function recompute is called in a special way. See man n trace.
trace variable a w {recompute}
```

# Message Sequence Chart

http://www.win.tue.nl/~michelr/mscintro/mscintro.html

Message Sequence Charts is a graphical and textual language for the description and specification of the interactions between system components and their environment. The language is standardized by the ITU-TS (the Telecommunication Standardization section of the International Telecommunication Union, the former CCITT).

The main area of application for Message Sequence Charts is as an overview specification of the communication behavior of real-time systems, in particular telecommunication switching systems. Message Sequence Charts may be used for requirement specification, interface specification, simulation and validation, test-case specification and documentation of real-time systems.

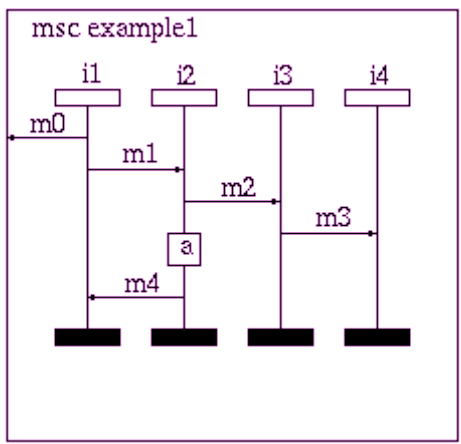 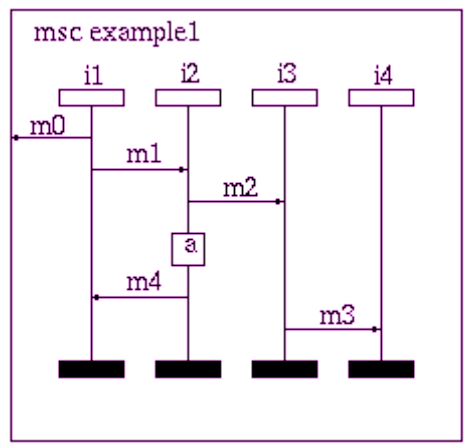

# State transition diagram

http://www.csc.calpoly.edu/~dbutler/tutorials/winter96/rose/node10.html

The State Transition diagram shows the state space of a given context, the events that cause a transition from one state to another, and the actions that result.

http://www4.informatik.uni-erlangen.de/Projects/JX/Projects/TCP/tcpstate.html

The TCP protocol state transition diagram

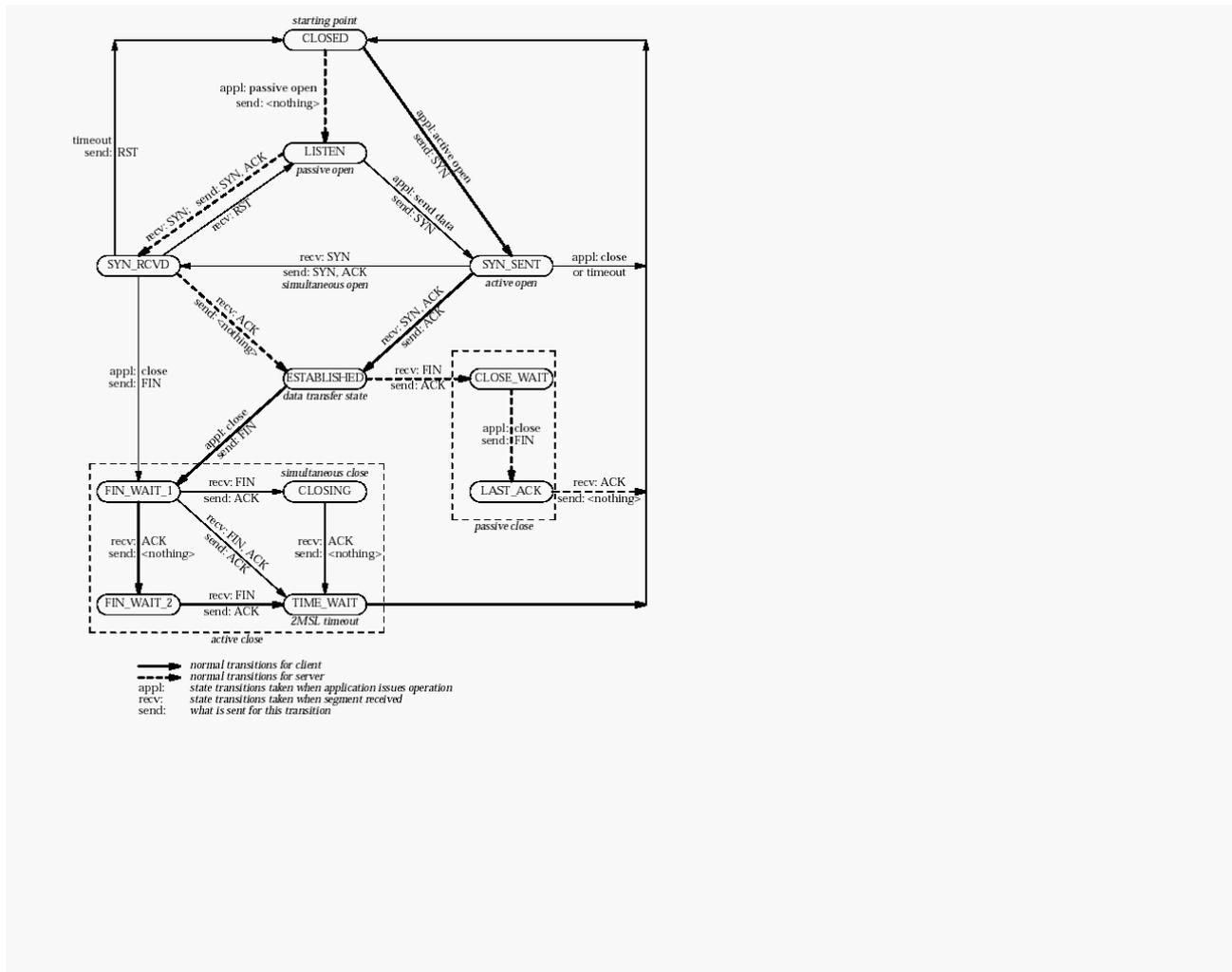

# System Architecture Steps

## *Spiral Model*

The spiral model is the over-all pattern (or paradigma or rule) to the system definition, design, implementation, test and roll-out. Using the scientific method as strategy of all steps in the spiral model the test (the experiment, the verification) is the most important one. To stress this point, we can talk of "test driven development"[8].
The test driven idea can be used for every phase in the spiral model. A requirement is only a good requirement if there is a test defined (or sketched) to test this requirement. If you can not think of a test, you have a wish but not a requirement is the conclusion.

## *Hierachical Decomposition and bottom-up development*

Hierachical decomposition is top-down planning. Every spiral model cycle should start with top-down planning. The implementation of a system is done best with a bottom-up approach. In every spiral model cycle the top-down planning leads to a bottom-up development that leads to an update of the hierachical decomposition plan.

Do a little planning, do a little work, do more planning, do more work is the mantra.

## *Model View Controller*

The MVC paradigma approaches system design from the (graphical) user interface, from the outside. To bring live into the user interface (and into the system) message sequence chart and state transition diagram are used. It is important for the following steps that all changes update the model. The model is the common element between the user-interface (the clock face) and the processing (the clock work) of the system which is done with algorithms, messages and state changes.

## *Message Sequence Chart and State Transition Diagram*

Often the behaviour of one detail of the system can be expressed as message sequence chart (MSC) or as state transition diagram (STD). See the TCP protocol state diagram for an example.
The MSC shows the time relation between the messages better then the STD. It is often easier to translate a use-case into a MSC.
The STD is better to define the error handling. Often the use-cases only cover the normal system behaviour, not the behaviour in case of message loss, message duplication or timeout.

# Conclusion

There is only one spiral model and only one Model-View-Controller paradigma. But hierachical decomposition and bottom-up development have a dualistic nature: one complements the other. This is very good to approximate the truth in the development process: approach the topic from more then one angle.
The same holds for message sequence chart and state transition diagram: The same information can be presented in both ways. And it makes a lot of sense to draw both charts: You get two views on the topic. This helps to draw a correct picture and build a correct system.

---

[8] http://en.wikipedia.org/wiki/Test_driven_development